\documentclass[reprint, showpacs, showkeys, superscriptaddress, aps, pre, twocolumn, 10pt]{revtex4-1}

\usepackage[T1]{fontenc}
\usepackage[utf8]{inputenc}
\usepackage{amsmath}
\usepackage{amssymb}
\usepackage{enumitem}
\usepackage{graphicx}
\usepackage{refstyle}
\usepackage{multirow}
\usepackage[normalem]{ulem}
\newcommand\redout{\bgroup\markoverwith
{\textcolor{red}{\rule[.5ex]{5pt}{0.5pt}}}\ULon}
\usepackage[
  citecolor=blue,
  colorlinks,
  linkcolor=blue,
  urlcolor=blue,
]{hyperref}
\usepackage[dvipsnames]{xcolor}

\newcommand{\be}{\begin{equation}}
\newcommand{\ee}{\end{equation}} 
\newcommand{\bea}{\begin{eqnarray}}
\newcommand{\eea}{\end{eqnarray}}

\begin{document}
\title{Replicator equations induced by microscopic processes in nonoverlapping population playing bimatrix games}
\author{Archan Mukhopadhyay}
\email{archan@iitk.ac.in}
\affiliation{
 Department of Physics,
  Indian Institute of Technology Kanpur,
  Uttar Pradesh 208016, India
}
\author{Sagar Chakraborty}
\email{sagarc@iitk.ac.in}
\affiliation{
  Department of Physics,
  Indian Institute of Technology Kanpur,
  Uttar Pradesh 208016, India
}

\begin{abstract}
This paper is concerned with exploring the microscopic basis for the discrete versions of the standard replicator equation and the adjusted replicator equation. To this end, we introduce frequency-dependent selection---as a result of competition fashioned by game-theoretic consideration---into the Wright--Fisher process, a stochastic birth-death process. The process is further considered to be active in a generation-wise nonoverlapping finite population where individuals play a two-strategy bimatrix population game. Subsequently, connections among the corresponding master equation, the Fokker--Planck equation, and the Langevin equation are exploited to arrive at the deterministic discrete replicator maps in the limit of infinite population size.
\end{abstract}
\maketitle
\maketitle
\section{Motivation}
The substantial literature on the replicator equations across disciplines epitomizes the pithy aphorism that ``All models are wrong, but some are useful'': In spite of being devoid of many realistic details of the respective systems, the replicator equations---the stylized  equations of evolutionary dynamics---have brought forth an insightful playground for understanding the theoretical aspects of the evolutionary game theory as studied in physics~\cite{hauert2005ajp, hidalgo2006pa,helbing2010pre,sadhukhan2020prr}, chemistry~\cite{stadler1991bs,stadler2003acs}, mathematics~\cite{shahshahani1979book}, social science~\cite{wang2009pre,montgomery2010aej,weitz2016pnas,lin2019prl,tilman2020nc}, economics~\cite{damme1994eer,weibull1994eer,samuelson1998book,friedman1998jee,newton2018games}, biology and ecology~\cite{hofbauer1998book,smith1982book,nowak2004science,tokita2004prl,liao2014if}, and reinforcement learning~\cite{borgers1997jet, sato2002pnas, sato2003pre, traulsen2004prl, galla2009prl, kianercy2013pre, jurgen2014pre, bloembergen2015jair}. Strictly speaking, the deterministic replicator equations, whether continuous or discrete in time, are models for the case of infinite-sized population of agents or individuals; but for a reasonably large enough population, the replicator dynamics are quantitatively decent approximations of the corresponding stochastic models~\cite{foster1990tpb,binmore1992book,boylan1992jet,cabrales1992jet,fudenberg1992jet,boylan1995jet,samuelson2002jep,fudenberg2015econometrica}.

Probably the most fundamental aspect of the replicator equations is the direct connection of their asymptotic dynamical states with the Nash equilibria of the normal form games' payoff matrices that are used in the equations. This fact is known as the folk theorem~\cite{nachbar1990ijgt,cressman2014pnas} of the evolutionary game theory. The theorem essentially asserts that a stable fixed point or a $\omega$-limit point of the replicator dynamics is a Nash equilibrium; and a strict Nash equilibrium corresponds to a asymptotically stable fixed point. The connection of the evolutionarily stable strategy/state with the fixed points of the replicator dynamics is less general; results differ depending on whether the equation is discrete or continuous. In this context, it should be borne in mind that it is rather unexplored~\cite{mukhopadhyay2020jtb} how the non-fixed point equilibrium sets (like periodic orbits and chaotic orbits) correspond to the game theoretic equilibria (like Nash equilibria or evolutionarily stable strategies).

It is natural to relate {the} discrete replicator equations with the populations having nonoverlapping generations and the continuous one with the overlapping generations case. But it is not uncommon to see {the} discrete replicator equation for the overlapping generations case~\cite{cabrales1992jet,binmore1992book}; in fact, there exists a discrete overlapping-generations replicator equation~\cite{weibull1997book} that spans the entire range of cases from strictly nonoverlapping case to overlapping (to different degrees) case. Nevertheless, {the} discrete replication equations that are not for generation-wise strictly nonoverlapping populations, in effect, means that the state of the population is observed after every finite interval of time---a factor that explicitly enters the discrete equations and when it goes to zero, yields a continuous replicator equation. A continuous differential equation can always be discretized in time (e.g., when solving it numerically) to yield a discrete difference equation that has the exact the same interpretation: It quantifies the evolution of the state of the system that is observed after a finite interval of time that can suitably be taken as the time interval between two successive generations. However, this exercise may mask the fundamental fact that a discrete equation is a model on its own; the profound importance of the logistic map~\cite{may1976nature}---\emph{vis-a-vis} its continuous version~\cite{piegton2004book}---in the development of the theory of chaos is there for all to see. 
\section{Introduction}
In line with the opening sentence of this paper, a very useful discrete replicator equation~\cite{borgers1997jet,bisin2000qje,hofbauer2000jee,montgomery2010aej,vilone2011prl,pandit2018chaos,mukhopadhyay2020jtb, mukhopadhyay2020chaos} is as follows:
\begin{equation}
x^{(k+1)}_i=x^{(k)}_i+ x^{(k)}_i[({\sf{\Pi}}{\bf x}^{(k)})_i-{\bf x}^{(k)T}{\sf{\Pi}} {\bf x}^{(k)}].
\label{eq:typeI_map}
\end{equation}
 It models the evolution of a population with $n$ different types and $x_i$ stands for the fraction of $i$th phenotype from the finite countable set of phenotypes.  ${\sf{\Pi}}$ is the payoff matrix of the corresponding population game and ${\bf x}$ is the state vector. The superscripts in brackets on $x$ denote the time step or generation number, and superscript $T$ represent the transpose operation. This is the simplest known replicator equation that is in compliance with the folk theorem, and simultaneously gives rise to periodic and chaotic orbits for the simple case of $2\times2$ normal form games~\cite{vilone2011prl,pandit2018chaos}. Therefore, it is a very useful stylized model that can be investigated to find the connections of the periodic and the chaotic orbits with the purely game theoretic concepts. Henceforth, for the sake of convenience, we call it type-I replicator map. Note that Eq.~(\ref{eq:typeI_map}) looks like the Euler forward discretization of the continuous (standard) replicator equation~\cite{taylor1978mb,schuster1983jtb,schuster1985bbpc,hofbauer1998book,page2002jtb,cressman2003book,traulsen2005prl},
\begin{equation}
\frac{d{x}_i}{dt}=x_i\left[({\sf{\Pi}}{\bf x})_i-{\bf x}^T{\sf{\Pi}} {\bf x}\right].
\label{eq:replicator_dynamics}
\end{equation}
The type-I replicator map (Eq.~(\ref{eq:typeI_map})) should be contrasted with the more popular discrete replicator equation, which we term type-II replicator map~\cite{taylor1978mb,dekel1992jet,hofbauer1998book,smith1982book,vanDamme1991book}, 
given below:
\begin{equation}
x^{{(k+1)}}_i=x^{(k)}_i\frac{({\sf{\Pi}}{\bf x}^{(k)})_i} {{\bf x}^{(k)T}{\sf{\Pi}} {\bf x}^{(k)}}.
\label{eq:typeII_replicator_map}
\end{equation}
The continuous dynamic, emanating from this discrete form, leads to the following equation:
\begin{equation}
\frac{d{x}_i}{dt}=x_i\frac{\left[({\sf {\Pi}}{\bf x})_i-{\bf x}^T{\sf{\Pi}} {\bf x}\right]}{{\bf x}^T{\sf{\Pi}} {\bf x}},
\label{eq:adjusted_replicator_dynamics}
\end{equation}
which is known as the adjusted replicator equation~\cite{smith1982book,hofbauer2000jee}. 

It is very well known~\cite{nachbar1990ijgt,weissing1991gem,cabrales1992jet,dekel1992jet,weibull1997book,bjornerstedt1997book} that particular results, e.g., stability and convergence in the temporal evolution, of a discrete replicator dynamic are not necessarily implied by the corresponding continuous replicator dynamic. The topologies of the solutions need not be same even if we take two different discrete replicator dynamics, viz., the type-I and the type-II replicator dynamics. We must recall~\cite{smith1982book} that topologically the solutions of both the continuous replicator equation and the adjusted replicator equation are equivalent for symmetric normal form games but it need not be so for asymmetric games with bimatrix normal form. Asymmetric games are frequently realized in natural scenarios, e.g., inter-species interactions, interactions between sub-populations, and intra-species interactions due to assignment of social roles~\cite{alex2015plos}. A few specific real life examples of inter-role bimatrix games are: competition between the owner and an invader of a territory~\cite{leimar1984jtb,grafen1987ab}, the cost associated with sex for a male and a female~\cite{clutton-brock1991book,jennions2017ptrsb}, and donation game between parent and offspring~\cite{marshall2009jtb}. Thus, the comparative cases of the corresponding discrete dynamics in the bimatrix games excites even more curiosity.

The microscopic basis of the evolutionary game theory relies on the probabilistic behavioural model~\cite{helbing1996td}. Both the continuous and the adjusted replicator equations have been shown to be the governing equations in the large population limit in a generation-wise overlapping population where a Moran process with respective selection mechanisms is in play~\cite{traulsen2005prl}. The question we are primarily asking in this paper is the following: Under what conditions can the discrete replicator equations (the type-I and the type-II replicator maps) be obtained as a large population limit of a generation-wise nonoverlapping population where the evolutionary dynamics is described by a microscopic stochastic birth-death process. In view of the above mentioned subtlety of the bimatrix games, we seek the answer to this question in the more general case of the bimatrix games.

Since it is clearly essential to start with a stochastic birth-death model that captures nonoverlapping generations, 
 our choice is to work with the seminal Wright--Fisher (WF) model~\cite{wright1931genetics, fisher1930book} that describes a biological population with strictly nonoverlapping generations. Here the reproduction process is synchronous, i.e., all reproduce at the same time. Next generation is chosen from the pool of these newly born offsprings. There are many real life examples of generation-wise non-overlapping populations of organisms such as periodical cicada~\cite{english2006ee}, annual plants~\cite{albani2010book}, pink salmon~\cite{rather2001ecology}, and squid~\cite{rather2001ecology}. The originally proposed WF model is devoid of any selection mechanism. Mathematical structure of the genetic drift in WF model is known in detail~\cite{tran2013tb}. Moreover, the statistical aspects of the WF process with a selection mechanism incorporated is also well-studied~\cite{imhof2006jmb,tataru2016sb}. 
 
The WF model has been successful in finding some of the statistical features of real life biological systems, e.g., the effective population size in molecular evolution~\cite{charlesworth2009nrg}, the population divergence time in chimpanzee~\cite{tataru2015genetics}, and the evolution of drug resistance of the influenza virus~\cite{ferrer-admetlla2016genetics}. Moreover, a recent paper~\cite{nelson2020plosgenetics} has shown that the WF process can model the genomic data of recent past quite accurately. These are remarkable practical applications of the WF model that, in essence, is a rather simple mathematical model of the birth-death process in a non-overlapping population.
  
In this paper, we show that starting with a selection driven WF process---appropriately modified to model two-strategy bimatrix  games---we can reach to the type-I and the type-II replicator maps as the governing deterministic dynamics in the infinite population limit. The WF process, being a discrete time Markov process~\cite{gardiner2004book,vankampen2007book}, facilitates the use of the standard links among the master equation, the Fokker--Planck equation, and the Langevin equation to arrive at the maps.

\section{The Model}
We consider a haploid population of size $N$ consisting of individuals with two types of roles denoted by  $\alpha$ and $\beta$. The entire collection of individuals of $\alpha$ role constitute a subpopulation that we term $\alpha$-population of size $N_\alpha$ and, remaining $N_\beta=N-N_\alpha$ individuals constitute $\beta$-population. Now, being confined to the one-locus-two-allele theory, the state of the entire population is given by the allele frequency at the locus under consideration; any individual (of either role) has one of the two alleles, either $A$ or $B$, at the locus. Hence, for the purpose of our calculations, an individual can be fully specified by an allele or (pheno-)type, and a role. 

We denote the state of $\theta$-population (here and henceforth, $\theta\in\{\alpha,\beta\}$) at generation $k$ as $i_{\theta}$ which stands for the number of alleles belonging to type $A$ and role $\theta$. Hence, the joint state of the system can be represented as $(i_{\alpha},i_{\beta})$. For $\theta$-population with a total of $N_{\theta}$  number of alleles, $i_{\theta}$ can have any value between $0$ and $N_{\theta}$. Let the probability of choosing type $A$ individual from $\theta$-population as a parent is given by $p_\theta$; consequently, the probability of choosing type $B$ is $1-p_{\theta}$. As in the case in the standard WF model, the transition probability $T^{\theta}_{i_{\theta},j_{\theta}}$ for $\theta$-population  from the current state $i_{\theta}$ to state $j_{\theta}$  in the next generation is taken as
\begin{eqnarray}
&&T^{\theta}_{i_{\theta},j_{\theta}}={{N_{\theta}}\choose{j_{\theta}}}\left(p_{\theta}\right)^{j_{\theta}}\left(1-p_{\theta}\right)^{N_{\theta}-j_{\theta}};\,\theta\in\{\alpha,\beta\}.
\label{eq:transition_probability_general form}
\end{eqnarray}
Thus, it is a microscopic stochastic evolutionary dynamics having synchronous reproduction where the offsprings in a generation of a subpopulation are chosen through a binomial random sampling with replacement from the immediately preceding generation of the same subpopulation. In the absence of any selection, i.e., for the case of pure genetic drift, $p_{\theta}=i_{\theta}/N_{\theta}$.

\subsection{Incorporation of selection}
\label{sec:ios}
In the presence of selection, the probability $p_\theta$ of choosing allele $A$ with role $\theta$ as parent must be dependent on the fitness of the individuals with role $\theta$. In the paradigm of natural selection, the fitness of a type is measured by the expected reproductive growth rate of that type~\cite{brown2016prsb}. There may be different ways in which $p_{\theta}$ may depend on the fitness profile of the whole population. Below we present two simplest {logical} possibilities:
\begin{enumerate}
\item In the most frequently used form~\cite{imhof2006jmb,TPI2006PRE}, $p_{\theta}$ is the ratio of effective reproductive fitness of all individuals belonging to allele $A$ of $\theta$-population to the effective reproductive fitness of the whole $\theta$-population:
\begin{eqnarray}
&& p_{\theta}=\frac{i_{\theta}f_{\theta}^{A}}{{i_{\theta}f_{\theta}^{A}}+(N_{\theta}-i_{\theta})f_{\theta}^{B}};\,\theta\in\{\alpha,\beta\}.
\label{eq:transition_probability_Type_II}
\end{eqnarray}
Here, $f_{\theta}^{A}$ and $f_{\theta}^{B}$ are respectively the fitnesses of type $A$ and type $B$ individual in $\theta$-population. It is clear from Eq.~(\ref{eq:transition_probability_Type_II}) that states $(i_{\alpha},i_{\beta})$, where $i_{\alpha}\in \{1,2,\cdots,N_{\alpha}-1\}$ and $i_{\beta}\in\{1,2,\cdots,N_{\beta}-1\}$, are transient states; and for $i_{\alpha}\in\{0,N_{\alpha}\}$ and $i_{\beta}\in\{0,N_{\beta}\}$, the $(i_{\alpha},i_{\beta})$ are absorbing states. If the initial state of the system is in any one of the transient states then it ultimately reaches to one of the absorbing states in finite time and stays there forever.

\item In the presence of selection, it is of higher probability that an offspring has a type $A$ individual as parent if the fitness of type $A$ is more than the fitness of type $B$.  This comparison between fitnesses should be done in a pair of randomly chosen individuals---drawn sequentially with replacement---when the pair consists of individuals of both types. The probability of choosing such a pair is $\left[{i_\theta}/{N_\theta}\right]\color{black}{\left[({N_\theta-i_\theta})/{N_\theta}\right]}$ and the comparison between fitnesses can be quantified by ${{f_{\theta}^{A}-{f}^B_{\theta}}}$. In conclusion, $p_{\theta}$ (which is only $i_\theta/N_\theta$ during exclusively random genetic drift) gets augmented by a contribution from the selection and mathematically it may be written down as follows to provide us with an alternate choice:
\begin{eqnarray}
&&p_{\theta}=\frac{i_{\theta}}{N_{\theta}}+\left(\frac{i_\theta}{N_\theta}\right)\left(\frac{N_\theta-i_\theta}{N_\theta}\right)\left(\frac{{f_{\theta}^{A}-{f}^B_{\theta}}}{\Delta f^{\rm max}_{\theta}}\right);
\label{eq:transition_probability_Type_I}
\end{eqnarray}
$ \theta\in\{\alpha,\beta\}$.
Here the positive proportionality constant ${\Delta f^{\rm max}_{\theta}}$ must be chosen in such a way that $p_{\theta}$ always stays non-negative and not more than unity. One possible choice is that ${\Delta f^{\rm max}_{\theta}}$ be the maximum possible value of absolute difference between the fitnesses of the two types in {the} $\theta$-population. It can be easily noted that the transient states and the absorbing states are same as the ones corresponding to Eq.~(\ref{eq:transition_probability_Type_II}). One can easily verify that if ${f_{\theta}^{A}={{f}}^B_{\theta}}$ for $\theta=\alpha$ and $\theta=\beta$, then we get back the standard WF model with no selection and only drift. This is true for the other case (Eq.~(\ref{eq:transition_probability_Type_II})) as well.
\end{enumerate}

The next question is how exactly one should quantify the fitnesses used above? This reproductive fitness is derived by comparing how one type with one specific role performs against the subpopulation of the other role when randomly matched. Independent of this interaction, there may exist a baseline fitness. Hence, an effective reproductive fitness of the individuals should be defined as,
\begin{subequations}
\begin{eqnarray}
&&f_{\theta}^{A}=1-w+w\pi_{\theta}^{A},\\
&&f_{\theta}^{B}=1-w+w\pi_{\theta}^{B},
\end{eqnarray}
\label{eq:fab}
\end{subequations}
where $\theta\in\{\alpha,\beta\}$ and $w$ is the strength of selection weighing the contributions, $\pi_{\theta}^{A}$ and $\pi_{\theta}^{B}$ (respectively for the type A and the type B individuals in $\theta$-population), owing to the inter-subpopulation interaction. Such an expression for fitness assumes that $w$ is the probability that the fitness (i.e., the expected growth rate) is solely derived from the inter-role interactions whereas $1-w$ is the probability that only the baseline fitness (fitness in the absence of any interaction) is contributing. For the purpose of simplicity, the baseline fitness is assumed to be independent of the type; by construction, it is same for both the types and has no role in selection. Hence, the contribution towards the selective advantages (which drive the natural selection process) are solely derived from the fitness term due to the inter-role interactions which is weighted by $w$. Hence, the choice of the phrase---strength of selection---to describe $w$ is apt.

\subsection{Frequency-dependent selection}

While the fitness of an individual of one subpopulation deceptively seems to be independent of the state of the other subpopulation, it should definitely be not so. The fitness must depend on the environment that an individual is in. All other individuals in the entire population effectively act as an environment. The evolutionary game theory is arguably an appropriate and efficient technique to grasp the essence of frequency dependent selection~\cite{brown2016prsb}. For a matrix game---having a finite number of strategies (or types)---each payoff element stands for the contribution towards the growth rate of a type resulting from a interaction corresponding to a particular strategy profile. Thus, we resort to the standard path of invoking game-theoretic consideration that naturally suits such a scenario.

The inter-subpopulation interactions, which we have considered up to now, are purely asymmetric in nature. The information about these interactions is contained in the following two-strategy bimatrix normal form game:
 \begin{eqnarray*}  
\centering
\begin{tabular}{cc|c|c|}
		& \multicolumn{1}{c}{} & \multicolumn{2}{c}{{Role $\beta$}}\\
		& \multicolumn{1}{c}{} & \multicolumn{1}{c}{$A$} & \multicolumn{1}{c}{\,\,\,\,$B$\,\,\,\,}\\\cline{3-4} 
		\multirow{2}*{Role $\alpha$}  & $A$ & $a_{\alpha},a_{\beta}$ & $b_{\alpha},c_{\beta}$ \\\cline{3-4}
		& $B$ & $c_{\alpha},b_{\beta}$ & $d_{\alpha},d_{\beta}$ \\\cline{3-4}
\end{tabular}
\end{eqnarray*}
where the first element and the second element in each box stand for the payoffs of an individual from $\alpha$-population and $\beta$-population respectively. The respective payoff matrices are more explicitly and compactly given below:
\begin{eqnarray}
{\sf \Pi}_\theta=
\left[\begin{matrix}
a_\theta & b_\theta \\
c_\theta & d_\theta
\end{matrix}\right];\,\theta\in\{\alpha,\beta\}.
\end{eqnarray}
If we assume that every individual interacts with a sample of individuals in the opposite role, then the average payoff of the individuals must be functions of the fractions of the two alleles in the other sub-population. In other words, if we consider the sample to be the entire sub-population in the opposite role, then the average payoff per interaction can be written down as,
\begin{subequations}
\begin{eqnarray}
\left[\begin{matrix}
\pi_{\alpha}^{A} \\
\pi_{\alpha}^{B}\end{matrix}\right]=
{\sf \Pi}_\alpha\left[\begin{matrix}
{i_\beta}/{N_\beta} \\
1-{i_\beta}/{N_\beta}\end{matrix}\right],\\
\left[\begin{matrix}
\pi_{\beta}^{A} \\
\pi_{\beta}^{B}\end{matrix}\right]=
{\sf \Pi}_\beta\left[\begin{matrix}
{i_\alpha}/{N_\alpha} \\
1-{i_\alpha}/{N_\alpha}\end{matrix}\right].
\end{eqnarray}
\label{eq:matrixpi}
\end{subequations}

In the light of these expressions, it is now explicit that the fitness of an individual (see Eqs.~(\ref{eq:fab})) of one sub-population is {completely} dependent on the state of the other sub-population as it should be. It is further clear that the strength of selection, $w$, decides the relative contribution between baseline fitness and the fitness from the game theoretic interactions~\cite{nowak2004nature}. One may note that, mathematically, the case $w<1$ is equivalent to the case $w=1$ with a different payoff matrix; the payoff elements of the resultant payoff matrix obviously depends both on the underlying game and the strength of selection $w$~\cite{claussen2005pre}. This expression of game-theoretic fitness has been successfully used to model many real life events such as the evolution of human intestinal microbiota~\cite{wu2016games} and chimpanzee choice rates in competitive games~\cite{martin2014sr}.

Equipped with the information about how {the} differential reproduction modelled by {the} differential fitnesses leads to (natural) selection mediated evolution, we now turn our attention {towards} writing the master equation for the (modified) WF process that is a discrete time Markov process. We are driven by the expectation that in the limit of infinite population, there must be a governing dynamics in discrete time that captures the essence of the evolutionary game under consideration. 
\section{Towards deterministic dynamics}
The master equation~\cite{gardiner2004book,vankampen2007book} of the WF process we are dealing with is a stochastic process and can be written follows:
\begin{equation}
P^{(k+1)}_{i_{\theta}}-P^{(k)}_{i_{\theta}}=\sum_{j_{\theta}=0}^{j_{\theta}=N_{\theta}}P^{(k)}_{j_{\theta}}T^{\theta}_{j_{\theta},i_{\theta}}-\sum_{j_{\theta}=0}^{j_{\theta}=N_{\theta}}P^{(k)}_{i_{\theta}}T^{\theta}_{i_{\theta},j_{\theta}}, 
\label{eq:master_equatioN_1}
\end{equation}
where $\theta \in \{\alpha,\beta\}$ and $P^{(k)}_{i_{\theta}}$ stands for the probability that the $\theta$-population is in {the} state $i_{\theta}$ at generation $k$. Subsequently, we rescale the system variables of the $\theta$-population to define {the} new variables: $x_{\theta}=i_{\theta}/N_{\theta}$, ${\tilde{x}}_{\theta}=j_{\theta}/N_{\theta}$, and $t_{\theta}={k}/N_{\theta}$. The corresponding probability density thus reads as $\rho_{\theta}(x_{\theta},t_{\theta})=N_{\theta}P^{(k)}_{i_{\theta}}$. Furthermore, for the convenience of mathematical manipulations, we write the rescaled transition matrix as a function of the state immediately before jump and the jump size: ${\widetilde T}^{\theta}({{{\tilde x}}_{\theta},r_{\theta}})=T^{\theta}_{N_{\theta}{{\tilde{x}}_{\theta}}, N_{\theta}x_{\theta}}$, where $r_{\theta}=x_{\theta}-{{\tilde{x}}_{\theta}}$. Thus, finally Eq.~(\ref{eq:master_equatioN_1}) can be expressed as follows: 
\begin{eqnarray}
\rho_{\theta}
(x_{\theta},t_{\theta}+N_{\theta}^{-1})
-\rho_{\theta}(x_{\theta},t_{\theta})=
\sum_{{r_{\theta}=x_{\theta}}}^{x_{\theta}-1}
\left[\rho_{\theta}(x_{\theta}-r_{\theta},t_{\theta})\right.\nonumber\\
\times\left.{\widetilde T}^{\theta}({x_{\theta}-r_{\theta},r_{\theta}})\right]-\sum_{{r_{\theta}=x_{\theta}}}^{x_{\theta}-1}\rho_{\theta}(x_{\theta},t_{\theta}){\widetilde T}^{\theta}({x_{\theta},-r_{\theta}}),\,\,\quad
\label{eq:master_equatioN_2}
\end{eqnarray}
where $\theta \in \{\alpha,\beta\}$. 

Progressing further with Eq.~(\ref{eq:master_equatioN_2}) is quite challenging~\cite{weber1986jap,tran2013tb,tran2015tb,tataru2016sb} because the transition matrix is not tridiagonal (e.g., the ones for the random walk and the Moran process) for the case of the binomial sampling. However, our primary aim in this paper being the derivation of the corresponding deterministic dynamic, the scope of our paper lies in the limit, $N_{\theta}\rightarrow\infty$. This is very fortunate because of the fact~\cite{papoulisbook} that the binomial distribution is rigorously approximated by a Gaussian distribution for any $p_\theta$ in this limit and the analytical tractability rendered by the Gaussian distribution is very convenient for {the} required mathematical manipulations. Thus, to being with, we approximate the transition matrix when $N_{\theta}\rightarrow\infty$ as follows:
\begin{eqnarray}
&&{\widetilde T}^{\theta}({{{\tilde x}}_{\theta},r_{\theta}})\rightarrow G(N_\theta{ x}_{\theta})=\frac{1}{\sqrt{2\pi \sigma_{\theta}^2 }}\exp\left[-\frac{(N_{\theta}{ x}_{\theta}-{\mu}_{\theta})^2}{2\sigma_{\theta}^2}\right],\nonumber\\
&&\Rightarrow{\widetilde T}^{\theta}({{{\tilde x}}_{\theta},r_{\theta}})\rightarrow \frac{1}{\sqrt{2\pi \sigma_{\theta}^2 }}\exp\left[-\frac{(N_{\theta}{ \tilde x}_{\theta}+N_{\theta}{ r}_{\theta}-{\mu}_{\theta})^2}{2\sigma_{\theta}^2}\right],\nonumber\\
&&\Rightarrow{\widetilde T}^{\theta}({{{x}}_{\theta},r_{\theta}})\rightarrow \frac{1}{\sqrt{2\pi \sigma_{\theta}^2 }}\exp\left[-\frac{(N_{\theta}{ x}_{\theta}+N_{\theta}{ r}_{\theta}-{\mu}_{\theta})^2}{2\sigma_{\theta}^2}\right]\nonumber\\
&&\phantom{{\widetilde T}^{\theta}({{{x}}_{\theta},r_{\theta}})\rightarrow }=G(N_\theta{ r}_{\theta}),
\label{eq:normal_dist}
\end{eqnarray}
 to the leading order as other higher order terms are comparatively negligible. Here, $G(N_\theta{x}_{\theta})$ is the probability density function of the Gaussian distribution in $N_\theta { x}_{\theta}$ with mean $\mu_{\theta}=N_{\theta}p_{\theta}$ and standard deviation $\sigma_{\theta}=\sqrt{N_{\theta}p_{\theta}(1-p_{\theta})}$. 
 From Eq.~(\ref{eq:normal_dist}) we note that for a fixed value of $x_\theta$, $G(N_\theta{x}_{\theta})$ can be equivalently seen as the Gaussian distribution ($G(N_\theta{r}_{\theta})$ being the probability density function) in $N_\theta r_{\theta}$ with mean $\mu_{\theta}-N_\theta { x}_\theta$ and standard deviation $\sigma_{\theta}$. 
 
With this in mind, in Eq.~(\ref{eq:master_equatioN_2}), we expand the probability densities in the Taylor series about $t_{\theta}$ in the left hand side and about $x_{\theta}$ in the right hand side; and similarly expand the transition probabilities about $x_{\theta}$ to finally arrive at the Fokker--Planck equation,
\begin{eqnarray}
&&\frac{\partial \rho_{\theta}}{\partial t_{\theta}}=-\frac{\partial}{\partial x_{\theta}}\left[\rho_{\theta}\left\{\lim_{N_\theta\rightarrow\infty}\int_{-N_\theta}^{+N_\theta}r_{\theta}G(N_{\theta}{r}_{\theta})d(N_{\theta}r_{\theta})\right\}\right]\nonumber\\
&&\phantom{\frac{\partial \rho_{\theta}}{\partial t_{\theta}}}+\frac{1}{2}\frac{\partial^2}{\partial x^2_{\theta}}\left[\rho_{\theta}\left\{\lim_{N_\theta\rightarrow\infty}\int_{-N_\theta}^{+N_\theta}r_{\theta}^2G(N_{\theta}{r}_{\theta})d(N_{\theta}r_{\theta})\right\}\right],\qquad
\label{eq:master_equation_4}
\end{eqnarray}
having ignored the higher order terms in $N_{\theta}^{-1}$. $\rho_{\theta}$ is shorthand for $\rho_{\theta}(x_{\theta},t_{\theta})$. Here we have made use of Eq.~(\ref{eq:normal_dist}) to replace ${\widetilde T}^{\theta}({{{x}}_{\theta},r_{\theta}})$ with its corresponding Gaussian approximation $G(N_{\theta}{ r}_{\theta})$ and consequently, the summations become integrals. The validity of such diffusion approximation to derive the Fokker--Planck equation is well investigated in population genetics~\cite{aalto1989jtb}. Assuming the stochastic noise to be uncorrelated in time, we can use the It$\hat{\rm o}$ calculus to reach the following form of the Langevin equation for $\theta \in \{\alpha,\beta\}$~\cite{ghbook, gardiner2004book,vankampen2007book, C2016C},
\begin{eqnarray}
&&{x}^{(k+1)}_{\theta}-{x}^{(k)}_{\theta}=\lim_{N_\theta\rightarrow\infty}\int_{-N_\theta}^{+N_\theta}r_{\theta}G(N_{\theta}{r}_{\theta})d(N_{\theta}r_{\theta})\nonumber \\
&&\phantom{{x}^{(k+1)}_{\theta}-{x}^{(k)}_{\theta}}+{\xi_{\theta}}\sqrt{\lim_{N_\theta\rightarrow\infty}\int_{-N_\theta}^{+N_\theta}r_{\theta}^2G(N_{\theta}{r}_{\theta})d(N_{\theta}r_{\theta})}~,\qquad
\label{eq:langevin_alpha}
\end{eqnarray}
where $\xi_{\theta}$ is the Gaussian white noise. While writing Eq.~(\ref{eq:langevin_alpha}), we have once again reminded ourselves that the WF model fashions {the} birth-death process in a population with nonoverlapping generations, and hence the discretized version of the Langevin equation is what we should use. The stochastic term has $1/\sqrt{N_{\theta}}$ dependance, implying that only the deterministic term contributes for {the} infinite population. Therefore, the discrete dynamics for the evolution of the states of $\theta$-population in the very large, generation-wise nonoverlapping population is given as
\begin{eqnarray}
{x^{(k+1)}_{\theta}}=p_{\theta}(x^{(k)}_{\alpha},x^{(k)}_{\beta});\,\theta \in \{\alpha,\beta\}.
\label{eq:langevin_3}
\end{eqnarray}
Here, considering Eqs.~(\ref{eq:transition_probability_Type_II})--(\ref{eq:matrixpi}), we have explicitly highlighted that $p_{\theta}$ is a function of both $x_{\alpha}$ and $x_{\beta}$.
 
Subsequently, in the light of Eq.~(\ref{eq:fab}), Eq.~(\ref{eq:matrixpi}) and Eq.~(\ref{eq:langevin_3}), the two different models of selection---viz., Eq.~(\ref{eq:transition_probability_Type_II}) and Eq.~(\ref{eq:transition_probability_Type_I}) as discussed in Sec.~\ref{sec:ios}---respectively yield
\begin{subequations}
\begin{eqnarray}
&&{x^{(k+1)}_{\alpha}}=x^{(k)}_{\alpha}\frac{\left({{\sf{\Pi}}}_{\alpha}{{\bf x}^{(k)}_{\beta}}\right)_1} {{{\bf x}^{(k)T}_{\alpha}}{{\sf{\Pi}}}_{\alpha} {{\bf x}^{(k)}_{\beta}}},\\
&&{x^{(k+1)}_{\beta}}={x^{(k)}_{\beta}} \frac{\left({{\sf{\Pi}}}_{\beta}{{\bf x}^{(k)}_{\alpha}}\right)_1} {{{\bf x}_{\beta}^{(k)T}}{{\sf{\Pi}}}_{\beta} {{\bf x}^{(k)}_{\alpha}}};
\end{eqnarray}
\label{eq:typeII_map_v2}
\end{subequations}
and
\begin{subequations}
\begin{eqnarray}
&&{x^{(k+1)}_{\alpha}}={x^{(k)}_{\alpha}}+\frac{{x^{(k)}_{\alpha}}}{{{\Delta{f}}^{\rm max}_{\alpha}}}{\left[({{\sf{\Pi}}}_{\alpha}{{\bf x}^{(k)}_{\beta}})_1-{\bf x}^{(k)T}_{\alpha}{{\sf{\Pi}}_{\alpha}} {{\bf x}^{(k)}_{\beta}}\right]},\label{eq:typeI_map_v2a}\nonumber \\
\\
&&{x^{(k+1)}_{\beta}}={x^{(k)}_{\beta}}+\frac{{x^{(k)}_{\beta}}}{{{\Delta{f}}^{\rm max}_{\beta}}}{\left[({{\sf{\Pi}}}_{\beta}{{\bf x}^{(k)}_{\alpha}})_1-{{\bf x}_{\beta}^{(k)T}}{{\sf{\Pi}}_{\beta}} {{\bf x}^{(k)}_{\alpha}}\right]}.\nonumber \\
\label{eq:typeI_map_v2b}
\end{eqnarray}
\label{eq:typeI_map_v2}
\end{subequations}
  \begin{figure}
	\centering
	\includegraphics[scale=0.44]{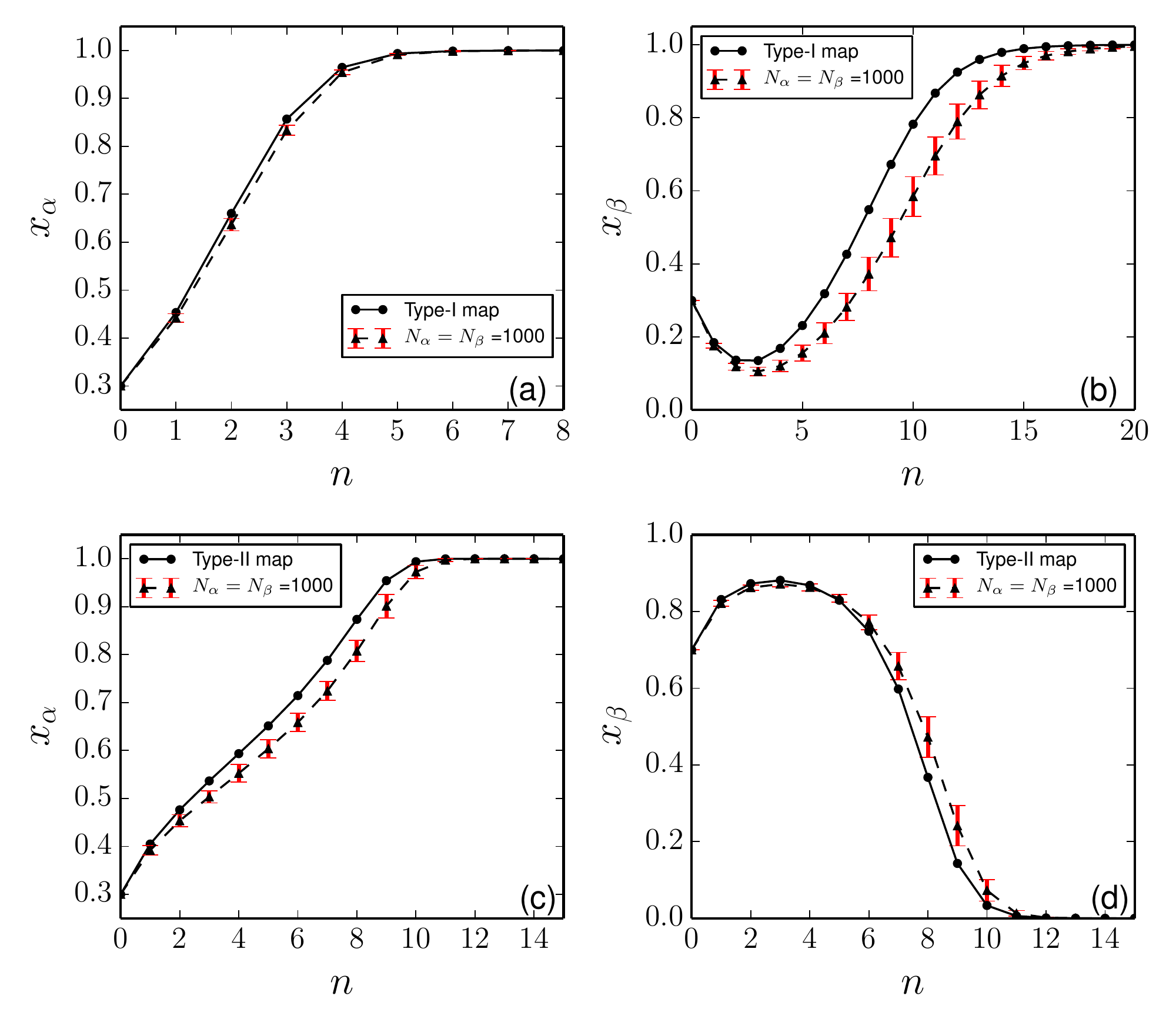} 		  
	\caption{Two dimensional replicator dynamics, type-I bimatrix dynamics (Eq.~(\ref{eq:typeI_map_v2})) and type-II bimatrix dynamics (Eq.~(\ref{eq:typeII_map_v2})), are mean field dynamics of the WF process in the very large population limit. Subplots (a) and (b) exhibit the dynamics of $\alpha$-population and $\beta$-population respectively; black solid lines are the evolution under Eq.~(\ref{eq:typeI_map_v2}) and black dashed lines with red error bars are the corresponding simulated WF process averaged over 50 trials for $N_{\alpha}=N_{\beta}=1000$. Subplots (c) and (d) are the analogous plots corresponding to the other case, viz., Eq.~(\ref{eq:typeII_map_v2}). For illustrative purposes, we consider that the $\alpha$-population is playing Harmony game ($a_{\alpha}=1$, $b_{\alpha}=1$, $c_{\alpha}=0.9$, and $d_{\alpha}=0$) in both the cases whereas the $\beta$-population is playing Prisoner's Dilemma game ($a_{\beta}=1$, $b_{\beta}=-1$, $c_{\beta}=0.5$, and $d_{\beta}=0$) and Leader game ($a_{\alpha}=1$, $b_{\alpha}=5$, $c_{\alpha}=6$ and $d_{\alpha}=0$) respectively in the cases corresponding to the type-I map and the type-II map. In subplots (a) and (b), the initial conditions chosen is $(x_{\alpha}=0.30$, $x_{\beta}=0.30)$ while in subplots (c) and (d)  $(x_{\alpha}=0.30$, $x_{\beta}=0.70)$ is chosen as the initial condition.}
	\label{fig_1}
\end{figure} 
The subscript, $1$, denotes the first row of the corresponding column vectors. Thus, we have successfully provided a microscopic basis for the discrete versions of the standard and the adjusted continuous replicator equations for the case of a two-strategy bimatrix population game in a generation-wise nonoverlapping population; we also validate our findings numerically in FIG.~\ref{fig_1}. 

\begin{figure}
	\centering
	\includegraphics[scale=0.44]{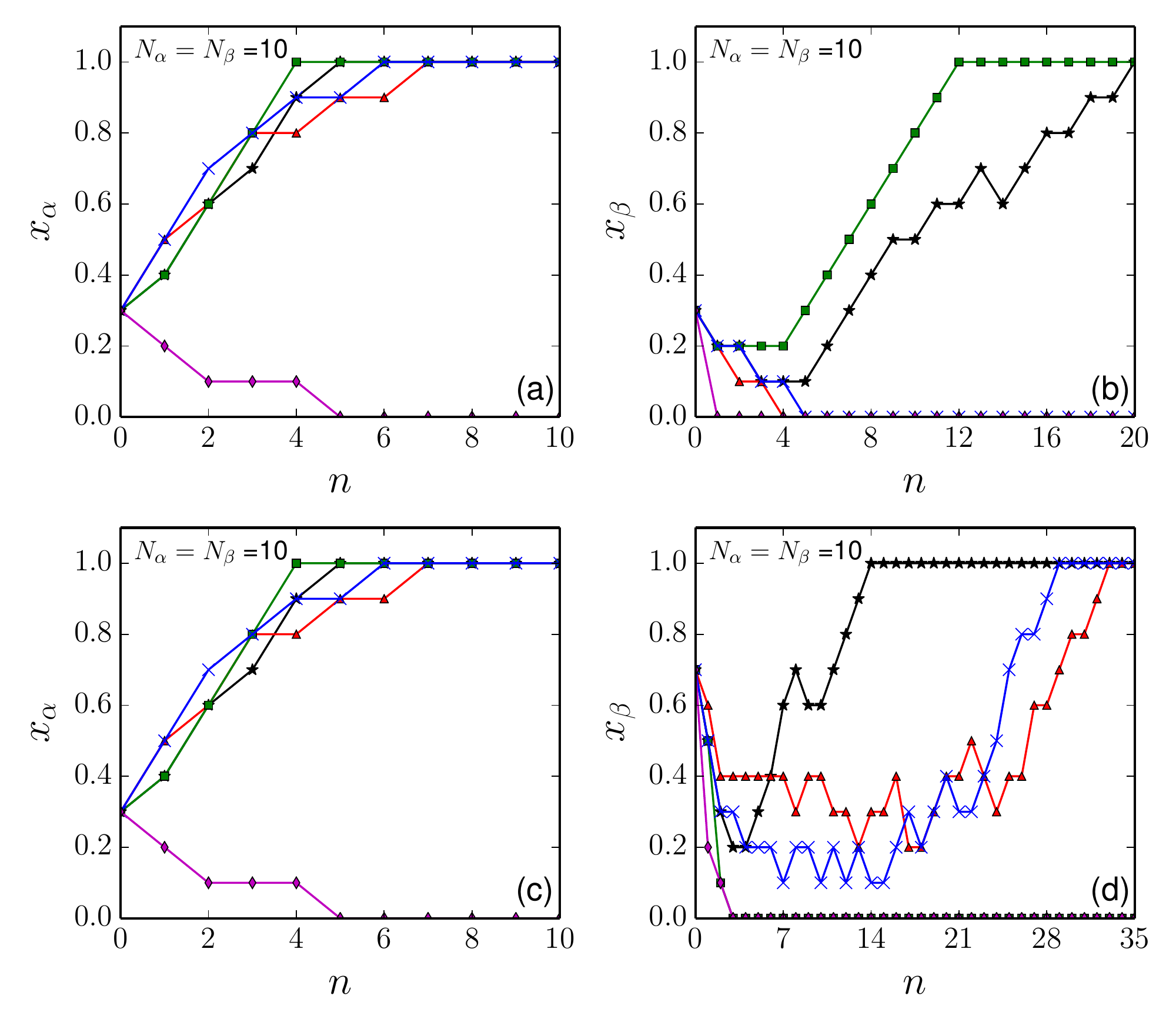} 		  
	\caption{For a finite population size, the effect of stochasticity leads the sub-population to randomly fixate in one of the two types, all $A$ or all $B$. Subplots (a) and (b) exhibit the dynamics of the $\alpha$-population and the $\beta$-population respectively for the same underlying games and initial conditions as used in FIG.~\ref{fig_1}(a) and FIG.~\ref{fig_1}(b); each coloured line is the simulated WF process generated in a random trial for $N_{\alpha}=N_{\beta}=10$. Similarly, subplots (c) and (d) exhibit the same scenario for the $\alpha$-population and the $\beta$-population respectively for the underlying games and initial conditions used in FIG.~\ref{fig_1}(c) and FIG.~\ref{fig_1}(d).}
	\label{fig_2}
\end{figure}

In passing, through FIG.~\ref{fig_2}, we emphasize once more that the replicator maps are good approximations of the governing dynamics for the selection driven WF processes only in the large population limit---note that the stochastic contribution is inversely proportional to the square root of the population size. The presence of stochasticity, in the finite population, leads the population to fixate in any one of its absorbing states---the sub-populations eventually fixate either into all $A$ or into all $B$ players---with a non-zero probability. However, it is a challenging task to find these fixation probabilities analytically for a selection driven WF process because of the non-tridiagonal form of the transition matrix. In future, it may be insightful to investigate if one can make use of the available approaches~\cite{ewens2004book, gardiner2004book, antal2006bmb, assaf2010jsm}---using either the discrete master equation or the corresponding Fokker--Plank equation---adopted for finding the fixation probability of a Moran process involving symmetric games (or even asymmetric games~\cite{sekiguchi2017dga}).

Coming back to Eqs.~(\ref{eq:typeI_map_v2a})--(\ref{eq:typeI_map_v2b}), please note the explicit presence of the positive definite quantities---${{\Delta{f}}^{\rm max}_{\alpha}}$ and ${{\Delta{f}}^{\rm max}_{\beta}}$--- that, at first glance, may be thought to be absorbable into the respective payoff matrices. However, there is an interesting caveat that we discuss now. It is straightforward to see that if we remove the concept of the roles in the population and allow for game-theoretic interaction of any two individuals in the resultant population of size $N(=N_\alpha+N_\beta)$ in the form of a $2\times2$ normal form game with payoff matrix, ${\sf \Pi}={\sf{\Pi}}^{\alpha}={\sf{\Pi}}^{\beta}$, we get back Eq.~(\ref{eq:typeI_map}) and Eq.~(\ref{eq:typeII_replicator_map}). (This leads to a one-dimensional dynamics in a simplex, $\Sigma^2$, in contrast with a two-dimensional dynamics in $\Sigma^2\times\Sigma^2$ accessed by the bimatrix game.) Specifically, we arrive at the type-I replicator map (cf. Eq.~(\ref{eq:typeI_map})),
\begin{equation}
x^{(k+1)}=x^{(k)}+ \frac{1}{{{\Delta{f}}^{\rm max}}}x^{(k)}[({\sf{\Pi}}{\bf x}^{(k)})_1-{\bf x}^{(k)T}{\sf{\Pi}} {\bf x}^{(k)}], 
\label{eq:typeI_map_symmetric}
\end{equation}
where ${\Delta f^{\rm max}}$ is the maximum possible value of absolute difference between the fitnesses of the two types in {the} population with no roles. This equation, in contrast to Eq.~(\ref{eq:typeI_map}), is incapable of showing any non-fixed point outcomes (like periodic orbits and chaotic orbits); furthermore, unlike Eq.~(\ref{eq:typeI_map}), any real payoff matrix can be used in Eq.~(\ref{eq:typeI_map_symmetric}) without violating the constraint $x\in[0,1]$ at any time. Details are presented in Appendix~\ref{appendix:B}.
\section{Discussion and Conclusion}
We emphasize that similar derivation~\cite{traulsen2005prl} of the standard and the adjusted continuous replicator equations from the microscopic birth-death process is based on the Moran model that models birth-death process in a generation-wise overlapping population; but for the generation-wise nonoverlapping population, fundamentally different birth-death process---e.g., the WF model---must be used. It is not obvious \emph{a priori} that the deterministic equations in the large population limit in the two birth-death processes should be connected. Consequently, in this paper, we have undertaken to analytically prove that the mean field dynamics that approximates the selection-driven evolution in a microscopic stochastic birth-death process in a population, where generations are nonoverlapping, can indeed be modelled by replicator maps. We have found that the type-I replicator map and the type-II replicator map, which are dynamically drastically different~\cite{pandit2018chaos} even in the simplest case of $2\times2$ normal form games, are the manifestations of the different ways in which individual fitness affects the selection in the birth-death process. 

It is interesting to note that the mean-field dynamics of a selection driven WF process is the discretized version of the mean-field dynamics corresponding to a selection driven Moran process~\cite{traulsen2005prl} although these two processes have two different microscopic justifications. The literature of these two microscopic processes has often highlighted this similarity; in fact, it is known that for the limit of infinite population many of the statistical and genealogical properties of these two microscopic processes are approximately same~\cite{tataru2016sb,bhaskar2009bioinformatics}. The model of selection---mediated through the choice of $p_{\theta}$---plays a crucial role in this aforementioned similarity of the mean-field dynamics. Note that the increment of the fraction of a type in the offspring pool of the WF process is similar to the excess probability that this type is chosen as an offspring in the corresponding Moran process~\cite{traulsen2005prl}. Thus, the reason behind the similarity in the mean-field dynamics for these two processes is the particular incorporation of selection in the neutral model. However, this similarity is only for the mean-filed dynamics---the stochastic components of the two processes behaves differently (see Appendix~\ref{appendix:C}). We remark that a different form of $p_\theta$ could give rise to an altogether different mean field dynamics that may be investigated in future; in this paper, our goal has been to find a microscopic model that leads to the widely used versions of the replicator maps.  


We remark that the WF model has the known advantage over the Moran model in allowing for diploidy in the population; in fact, one can easily redo the entire exercise of this paper by considering a monoecious diploid population to arrive at the same replicator maps. It must be explicitly highlighted that instead of type-I equation given by Eq.~(\ref{eq:typeI_map}), the above mentioned microscopic process yields a slightly different Eq.~(\ref{eq:typeI_map_symmetric}) that is very attractive as it always gives rise to convergent fixed-point outcomes. From a different perspective, it also means that the problem of deriving (from some microscopic birth-death process) the type-I equation given by Eq.~(\ref{eq:typeI_map}) which can yield chaotic outcomes~\cite{vilone2011prl,pandit2018chaos} remains an open question. From our present study, it appears that one needs to go beyond both the Moran and the WF models to solve this problem.

While relatively less in vogue in the biological systems, Eq.~(\ref{eq:typeI_map})---as itself or in related forms---also appears in modelling intergenerational cultural transmission~\cite{bisin2000qje,montgomery2010aej}, boundedly rational players' imitational behaviour in bimatrix cyclic games~\cite{hofbauer2000jee}, and reinforcement learning~\cite{borgers1997jet}. It is interesting to recall that a good behaviour rule does not require aggregate population behaviour implicit in the natural selection to induce the replicator map; instead the map is arrived at based on the rational behaviour of the players~\cite{cressman2003book}. In such contexts, the `strategies' $A$ and $B$ are not hardwired in the individuals and they are free to choose strategies. In such a scenario, we can put $w=1$, i.e., the interactions are only game-theoretic and note that the selection process in the type-I replicator map (in contrast with the type-II replicator map) intriguingly requires the underlying microscopic interactions to depend only on location information (see Eq.~\ref{eq:transition_probability_Type_I}): At every time step a pair of randomly chosen individuals with opposite strategies compare each other's expected payoffs; in other words, the selection depends on $f^A_\theta-f^B_\theta=\pi^A_\theta-\pi^B_\theta$. (The other selection model requires the knowledge of the global information for the calculation of the average payoff; see the denominator of the right hand side of Eq.~(\ref{eq:transition_probability_Type_II}).)

Some of the basic driving forces of evolution are selection, mutation, drift, migration, and recombination. All the replicator equations discussed herein are concerned with selection driven evolution. The deterministic replicator equations, by construction, can not capture the effect of drift. Migration is best modelled into the equations after imparting some spatial structure in the system. Extensions of the replicator equations to incorporate the effect of mutation has been extensively researched~\cite{schuster1977naturwissenschaften,hadeler1981jam,stadler1992jmb,bomze1995geb,page2002jtb,kamarova2004jtb,eigen2007book,mittal2020pre}; mathematically, one can modify the fitness functions and derive different replicator-mutator maps in a generation-wise nonoverlapping population of haploid (or monoecious diploid) individuals (see Appendix~\ref{appendix:A}). The mechanism of recombination, and related effects, are beyond the scope of the present investigation; but they could be a potential avenue of future research in the context of this paper. 

It is interesting to note that while the two different selection mechanisms leads to two different replicator maps in the infinite population limit, some of their effects on the finite population are identical. For example, consider the simpler case of the symmetric normal form $2\times2$ games with no different roles: If the selection favours type $A$ individual in invading a population with type $B$ individuals and opposes the opposite scenario, then for every possible transient initial state, the probability that $A$ fixates is more than for the case where only drift (no selection) is in action. This follows~\cite{imhof2006jmb} from the fact that the submatrix formed by deleting the first and the last columns and rows of the transition matrix (corresponding to any of the two selection mechanisms) is positive-definite, and its every row-sum---consequently, the spectral radius---is less than unity. Similarly, one may also be interested in comparing between the two update rules, in terms of the stochastic gains due to the intrinsic noise arising as a result of finiteness of the population size~\cite{rohl2008pre}. More exhaustive studies on the comparison between the two selection mechanisms for the case of finite population may be worth pursuing in future.

Before we end, let us remind ourselves that which replicator equation is {a} better model of evolution is an issue open for debate. One of the earliest hint about this non-unanimity may be found in the seminal book by Maynard-Smith~\cite{smith1982book} where it is discussed whether the adjusted replicator dynamics is a better model than the standard replicator dynamics, given the fact that these two continuous dynamics cease to have topologically identical solutions for the asymmetric games. However, one must take this non-unanimity positively and constructively since the replicator equations are merely minimal models that easily incorporate the Darwinian tenet of natural selection to investigate mathematical system mimicking exclusively the replication-selection aspect of the real evolutionary systems. In this spirit, what we have achieved in this paper is to provide a more sound (microscopic) footing to the type I replicator map that has immense potential to serve as a convenient testbed for investigating the connections~\cite{pandit2018chaos,mukhopadhyay2020jtb} between the complex dynamical outcomes and game-theoretic equilibria. 
\acknowledgements
The authors are grateful to Jayanta K. Bhattacharjee, Debashish Chowdhury and Samrat Sohel Mondal for helpful discussions.
\section*{AIP Publishing data sharing policy} 
\textcolor{black}{The data that support the findings of this study are available from the corresponding author upon reasonable request.}
\appendix
\section{No Non-Fixed Point Outcomes in Eq.~(\ref{eq:typeI_map_symmetric})}
\label{appendix:B}
For the purpose of this section and without any loss of generality, it is sufficient to work with the following form of the payoff matrix~\cite{pandit2018chaos}: $\Pi= (\begin{smallmatrix} 1&S\\ T&0 \end{smallmatrix}$); $S,T\in\mathbb{R}$. For further convenience, we explicitly define a function $g(x^{(k)})$ as follows to rewrite Eq.~(\ref{eq:typeI_map_symmetric}) as
\begin{equation}
x^{{k+1}}=g(x^{(k)})= x^{(k)}+\frac{1}{2}H_{x^{(k)}}\left( \frac{Cx^{(k)}+S}{{{\Delta{f}}^{\rm max}}}\right),
\label{eq:typeI_map_symmetric_ST}
\end{equation}
where $C=1-T-S$ and $H_{x^{(k)}}=2x^{(k)}(1-x^{(k)})$.
One may note that, by definition,
\begin{eqnarray}
{{\Delta{f}}^{\rm max}}=\max_{x}|Cx+S|=\max\{|S|,|1-T|\}.\label{eq:TORM}
\end{eqnarray} 
We immediately note that for all $S,T\in\mathbb{R}$ in Eq.~(\ref{eq:typeI_map_symmetric_ST}), $x^{(k)} \in [0,1]$ for all $k \in \mathbb{N}\cup\{0\}$ because $0\le|(Cx^{(k)}+S)/{{\Delta{f}}^{\rm max}}| \le 1$ and $0\le H_{x^{(k)}}/2 \le (1-x^{(k)})\le1$ for all $x^{(k)} \in [0,1]$ and for all $S,T\in\mathbb{R}$.

Now to prove that no periodic orbit is possible in Eq.~(\ref{eq:typeI_map_symmetric_ST}), we only have to show that $g(x)-x$ and $g^m(x)-x$ ($m\in\mathbb{N}$), where $g^m(x)$ is the $m$th iterate of $g(x)$, have same sign for all $x\in[0,1]\backslash\mathbb{F}$; here, $\mathbb{F}$ is the set of all fixed points or 1-period points, i.e., the solutions of $g(x)=x$. (Note that, in the immediately preceding paragraph, we have already shown that $g^m(x)\in[0,1]$ for all $x\in[0,1]$.) This is so because it means that in the plot of $g(x)$ versus $x$, the number of intersections made with the line $g(x)=x$ is same as the number of intersections made between the line and the plot of $g^m(x)$ versus $x$; in other words, no prime $m$-period point (no new intersection) can appear. 

To this end, we first define $\delta_m=g^m({x})-x$ and use Eq.~(\ref{eq:typeI_map_symmetric_ST}) to find
\begin{equation}
\delta_{m+1}=
\delta_m\left(
1+\frac{1}{2}H_{g^m(x)}\frac{C}{{{\Delta{f}}^{\rm max}}}
\right)+\delta_1\frac{H_{g^m(x)}}{H_x}, 
\label{eq:typeI_map_symmetric_ST_delta_b}
\end{equation}
where $g^0(x)=x$.
Note that for the case of $m=1$,  Eq.~(\ref{eq:typeI_map_symmetric_ST_delta_b}) can be simplified as,
\begin{equation}
\delta_2=
\delta_1\left(
1+\frac{1}{2}H_{g^1(x)}\frac{C}{{{\Delta{f}}^{\rm max}}}
+\frac{H_{g^1(x)}}{H_x}\right).
\label{eq:typeI_map_symmetric_ST_delta_c}
\end{equation}
We have already shown that the map, expressed by Eq.~(\ref{eq:typeI_map_symmetric_ST}), is forward invariant: $g^{m}(x)\in[0,1]$ for all $m\in\mathbb{N}$. By definition, $0\le H_{g^m(x)} \le1/2$. If $C \ge 0$, it is trivial to see that the term in the parentheses of Eq.~(\ref{eq:typeI_map_symmetric_ST_delta_c}) is always positive which implies that $\delta_2$ and $\delta_1$ have same sign. Whenever $C <0$, it is straightforward to see that $|C/{{\Delta{f}}^{\rm max}}| < 2$ and hence, $(1/2)H_{g^m(x)}|C/{{\Delta{f}}^{\rm max}}|<1/2$ for all $m\in\mathbb{N}$. Therefore, $\delta_1$ and $\delta_2$ have same signs for all $x\in[0,1]\backslash\mathbb{F}$.

Following the similar arguments, it is easy to see from Eq.~(\ref{eq:typeI_map_symmetric_ST_delta_b}) that $\delta_1$ and $\delta_{m+1}$ have same sign if $\delta_1$ and $\delta_m$ have same sign. In the preceding paragraph, we have shown that $\delta_1$ and $\delta_2$ have same sign, and so from Eq.~(\ref{eq:typeI_map_symmetric_ST_delta_b})---with $m=2$---we conclude that $\delta_3$ and $\delta_1$ have same sign. Thus, using the method of induction, one concludes that $\delta_1$ and $\delta_m$ have same sign for all $m\in\mathbb{N}$ and no prime $m$-period points are possible. Furthermore, since there is no periodic orbit in the system, no chaotic attractor---that is supposed to have countably infinite number of unstable periodic orbits---can be present in the system.

\section{A brief note on the stochastic dynamics}
\label{appendix:C}
The master equation captures the time evolution of the probability distribution of a system and has the full information of the stochastic evolution of the model under consideration. Thus, in principle, one should be able to use it to derive the replicator map directly from the master equation~\cite{lin2019prl}. To this end, we multiply Eq.~(\ref{eq:master_equatioN_1}) by $i_\theta$ and sum over its all possible values to write the following:
\begin{eqnarray}
&&\sum_{i_{\theta}=0}^{i_{\theta}=N_{\theta}}{i_{\theta}} P^{(k+1)}_{i_{\theta}}- \sum_{i_{\theta}=0}^{i_{\theta}=N_{\theta}}{i_{\theta}} P^{(k)}_{i_{\theta}}=\sum_{j_{\theta}=0}^{j_{\theta}=N_{\theta}} P^{(k)}_{j_{\theta}}\sum_{i_{\theta}=0}^{i_{\theta}=N_{\theta}} {i_{\theta}} T^{\theta}_{j_{\theta},i_{\theta}}\nonumber \\
&&\phantom{\sum_{i_{\theta}=0}^{i_{\theta}=N_{\theta}}{i_{\theta}} P^{(k+1)}_{i_{\theta}}- \sum_{i_{\theta}=0}^{i_{\theta}=N_{\theta}}{i_{\theta}} P^{(k)}_{i_{\theta}}=}-\sum_{i_{\theta}=0}^{i_{\theta}=N_{\theta}} {i_{\theta}} P^{(k)}_{i_{\theta}} \sum_{j_{\theta}=0}^{j_{\theta}=N_{\theta}}  T^{\theta}_{i_{\theta},j_{\theta}}.\qquad
\label{4.alternative_1}
\end{eqnarray}
On using the form of transition matrix elements given by Eq.~(\ref{eq:transition_probability_general form}) and the properties of a binomial distribution, we find that $\sum_{i_{\theta}=0}^{i_{\theta}=N_{\theta}} {i_{\theta}}T^{\theta}_{j_{\theta},i_{\theta}}=N_{\theta}p_{\theta}(x^{(k)}_{\alpha},x^{(k)}_{\beta})$ and $\sum_{j_{\theta}=0}^{j_{\theta}=N_{\theta}}  T^{\theta}_{i_{\theta},j_{\theta}}=1$. Using these relations, we rewrite Eq.~(\ref{4.alternative_1}) as,
\begin{equation}
{\langle x_{\theta}^{(k+1)} \rangle}={\langle p_{\theta}(x^{(k)}_{\alpha},x^{(k)}_{\beta}) \rangle},
\label{4.alternative_2}
\end{equation}
where the pair of angular brackets denotes the ensemble average (average over the probability distribution). For the case of infinite population size ($N_{\theta} \to \infty$) we can safely approximate that the fluctuation around the mean is negligible. Hence, we can remove the angular brackets used in Eq.~(\ref{4.alternative_2}) to reach Eq.~(\ref{eq:langevin_3}) which further leads to Eq.~(\ref{eq:typeII_map_v2}) and Eq.~(\ref{eq:typeI_map_v2}) depending on the choice of $p_{\theta}(x^{(k)}_{\alpha},x^{(k)}_{\beta})$. We observe once more that Eq.~(\ref{4.alternative_2}), being the mean field dynamics, can not explicitly reveal the implications of the stochastic effects that is present in the system.

However, in order to find the stochastic component of the dynamics analytically, one must resort to further approximations. We note that the master equation under consideration has time-independent transition rates and hence the system it is describing is Markovian in nature. Consequently, its Kramer--Moyal expansion up to second order (an approximation leading to the Fokker--Planck equation) corresponds to the Langevin equation with a Gaussian delta-correlated noise~\cite{miguel2000book}. Of course, it would be more realistic to work ab initio with a Langevin dynamics having a coloured noise with a finite correlation time; however, not only the corresponding non-Markovian dynamics is generally analytically intractable but also it means that the corresponding master equation has time-dependent transition rates~\cite{haunggi1994book} at odds with the basic premise of the ideal Wright--Fisher process considered in this paper.

Within the approximation of the Markov process---modelled by the Langevin equation with white noise $\xi_{\theta}$---it is easy to find the stochastic component analytically: Calculations simplify Eq.~(\ref{eq:langevin_alpha}) to
\begin{eqnarray}
&&{x^{(k+1)}_{\theta}}=p_{\theta}(x^{(k)}_{\alpha},x^{(k)}_{\beta})+\sqrt{\frac{p_{\theta}(x^{(k)}_{\alpha},x^{(k)}_{\beta})[1-p_{\theta}(x^{(k)}_{\alpha},x^{(k)}_{\beta})]}{N_{\theta}}} {\xi_{\theta}}.\nonumber\\
\label{4.alternative_3}
\end{eqnarray}
 This clearly indicates that only the mean field dynamics (deterministic part) of the selection driven Moran process and the selection driven WF process are similar while the stochastic parts can be different, e.g., the amplitude of the stochastic contribution for a selection driven Moran process with local update rule (which leads to the replicator dynamics as the mean field dynamics) only depends on the state of the system~\cite{traulsen2005prl} whereas the same for the corresponding WF process (which has the type-I map as the mean-field dynamics) depends both on the state of the system and the underlying game (see Eq.~(\ref{4.alternative_3})).

\section{Replicator-Mutator Maps}
\label{appendix:A}
Within the scope of the replicator equations, the mutation effectively means that a particular type of individual reproduces the other type of individual. Let the probability that some $A$ type offsprings are born from an $A$ type individual is given by $q_{\theta}$ ($0\le q_{\theta}\le1$) in the $\theta$-population leaving the possibility that with probability 
$1-q_{\theta}$, some $B$ type offsprings are reproduced by an $A$ type individual. In general, in the case of a population with $n$ different types of individual, the probability that some of $j$th type individuals are reproduced by an $i$th type individual is given by $q_{ij}$ ($i,j\in\{1,2,\cdots,n\}$)---$n\times n$ elements of the mutation matrix that is row stochastic. For the $\theta$-population, further assuming that the mutation is symmetric between the $A$ and the $B$ type individuals, we note that the single parameter $q_{\theta}$ captures the mutation process. The resulting replicator-mutator map can be obtained by modifying the expressions of the fitnesses as follows:%
\begin{subequations}
\begin{eqnarray}
&&f_{\theta}^{A}=1-w+w\left[q_{\theta}\pi_{\theta}^{A}+(1-q_{\theta})\pi_{\theta}^{B}\right],\\
&&f_{\theta}^{B}=1-w+w\left[q_{\theta}\pi_{\theta}^{B}+(1-q_{\theta})\pi_{\theta}^{A}\right].
\end{eqnarray}
\end{subequations}
Following the method detailed in the text, the corresponding replicator-mutator maps, viz., type I and type II, for the two-strategy bimatix games are obtained respectively as:
\begin{subequations}
\begin{eqnarray}
&&{x^{(k+1)}_{\alpha}}={x^{(k)}_{\alpha}}+q_\alpha\left({{\sf{\Pi}}}_{\alpha}{{\bf x}^{(k)}_{\beta}}\right)_1{x^{(k)}_{\alpha}}+(1-q_\alpha)\left({{\sf{\Pi}}}_{\alpha}{{\bf x}^{(k)}_{\beta}}\right)_2\nonumber\\
&&\phantom{{x^{(k+1)}_{\alpha}}=}\times (1-{x^{(k)}_{\alpha}})-{x^{(k)}_{\alpha}}\left[{\bf x}^{(k)T}_{\alpha}{{\sf{\Pi}}_{\alpha}} {{\bf x}^{(k)}_{\beta}}\right],\\
&&{x^{(k+1)}_{\beta}}={x^{(k)}_{\beta}}+q_\beta\left({{\sf{\Pi}}}_{\beta}{{\bf x}^{(k)}_{\alpha}}\right)_1{x^{(k)}_{\beta}}+(1-q_\beta)\left({{\sf{\Pi}}}_{\beta}{{\bf x}^{(k)}_{\alpha}}\right)_2\nonumber\\
&&\phantom{{x^{(k+1)}_{\beta}}=}\times (1-{x^{(k)}_{\beta}})-{x^{(k)}_{\beta}}\left[{{\bf x}_{\beta}^{(k)T}}{{\sf{\Pi}}_{\beta}} {{\bf x}^{(k)}_{\alpha}}\right];
\end{eqnarray}
\label{eq:typeI_map_v2q}
\end{subequations}
and
\begin{subequations}
\begin{eqnarray}
&&{x^{(k+1)}_{\alpha}}=\frac{q_\alpha\left({{\sf{\Pi}}}_{\alpha}{{\bf x}^{(k)}_{\beta}}\right)_1x^{(k)}_{\alpha}+(1-q_\alpha)\left({{\sf{\Pi}}}_{\alpha}{{\bf x}^{(k)}_{\beta}}\right)_2(1-x^{(k)}_{\alpha})} {{{\bf x}^{(k)T}_{\alpha}}{{\sf{\Pi}}}_{\alpha} {{\bf x}^{(k)}_{\beta}}},\quad\nonumber\\
\\
&&{x^{(k+1)}_{\beta}}=\frac{q_\beta\left({{\sf{\Pi}}}_{\beta}{{\bf x}^{(k)}_{\alpha}}\right)_1{x^{(k)}_{\beta}} +(1-q_\beta)\left({{\sf{\Pi}}}_{\beta}{{\bf x}^{(k)}_{\alpha}}\right)_2(1-{x^{(k)}_{\beta}}) } {{{\bf x}_{\beta}^{(k)T}}{{\sf{\Pi}}}_{\beta} {{\bf x}^{(k)}_{\alpha}}}.\nonumber\\
\label{eq:typeII_map_v2q}
\end{eqnarray}
\end{subequations}
\bibliography{Mukhopadhyay_etal_manuscript.bib}
 \end{document}